\documentclass[fleqn,twocolumn,showpacs,showpreprint,amsmath,prd,amssymb]{revtex4}
\usepackage{graphicx}
\usepackage{psfrag}
\newcommand{\figwidth}{3.3in}
\begin{document}
\title{Direct WIMP Detection Using Scintillation Time Discrimination in Liquid Argon}
\newcommand{\lanl}{Physics Division, MS H803, Los Alamos National Laboratory, Los Alamos, NM 87545}
\author{M.G.~Boulay}\email[Corresponding author, email: ]{mboulay@lanl.gov}
\author{A.~Hime}\affiliation{\lanl}
\begin{abstract}
Discrimination between electron and nuclear recoil events in a liquid argon scintillation detector has been
demonstrated with simulations by using the differences in the scintillation photon time distribution between these classes of events.  A discrimination power greater than $10^{8}$ is predicted for a liquid argon experiment with a 10 keV threshold, which would mitigate electron and $\gamma$-ray backgrounds, including $\beta$ decays of ${}^{39}$Ar and ${}^{42}$Ar in atmospheric argon.   A dark matter search using a $\sim$ 2 kg argon target viewed by immersed photomultiplier tubes would allow a sensitivity to a spin-independent WIMP-nucleon cross-section
of $\sim10^{-43}$ cm$^{2}$ for a 100 GeV WIMP, assuming a one-year exposure.  This technique could be used to scale the target mass to the tonne scale, allowing a sensitivity of $\sim10^{-46}$ cm$^{2}$.
 \end{abstract}
\date{\today}
\maketitle

It is now well established that a significant part of the matter present in our universe is non-baryonic dark matter.
A currently favored hypothesis is that the dark matter is comprised  of WIMPs, Weakly Interacting Massive Particles,
which have so far remained undetected.  Direct searches place limits on the interaction cross-section and mass
of these particles, with the current best limit from the Cryogenic Dark Matter Search (CDMS) experiment~\cite{cdms}.  Scaling up the  current-generation experiments to large masses (on the order of a tonne) is of current interest and there are several proposed methods for achieving this goal.
Much effort has been placed on the study of noble liquids as detectors for WIMP particles in the past few years, since
many of these have high photon scintillation yields, can be cleaned to very low levels of radioactivity, and can 
be used to provide a large target mass for WIMP searches.
A large part of the experimental effort is in separating the ubiquitous backgrounds from $\gamma$ rays and electrons
which enter the sensitive detector region.
The XENON~\cite{XENON}, ZEPLIN~\cite{zeplin} and WARP~\cite{warp} collaborations are attempting to discriminate against
these backgrounds by drifting and subsequently detecting the energetic electrons which result from ionization in the detector, using liquid xenon and argon targets.  It has recently been shown that differences
in scintillation photon time distributions between electron and nuclear recoils could be exploited for effective background
discrimination in liquid neon~\cite{neon_sim}.  This discrimination capability, together with the potential to purify liquid neon using conventional cold traps, makes possible a dual purpose detector for low-energy solar neutrinos and WIMP dark matter.  Final evaluation of the discrimination achievable in neon awaits accurate determination of the scintillation parameters.  
In this paper, we follow the approach detailed in Ref.~\cite{neon_sim} and realize the opportunity to exploit scintillation time discrimination to reject backgrounds in a dedicated WIMP search using a liquid argon target.  
The small difference between the singlet and triplet time constants in liquid xenon precludes discrimination at the levels achievable in neon and argon based on timing alone. Consequently, a dedicated WIMP detector based on liquid argon can be realized that is conceptually simple and that does not require the complication associated with high-field drift of electrons.

\begin{figure}
\begin{center}
\includegraphics[width=\figwidth]{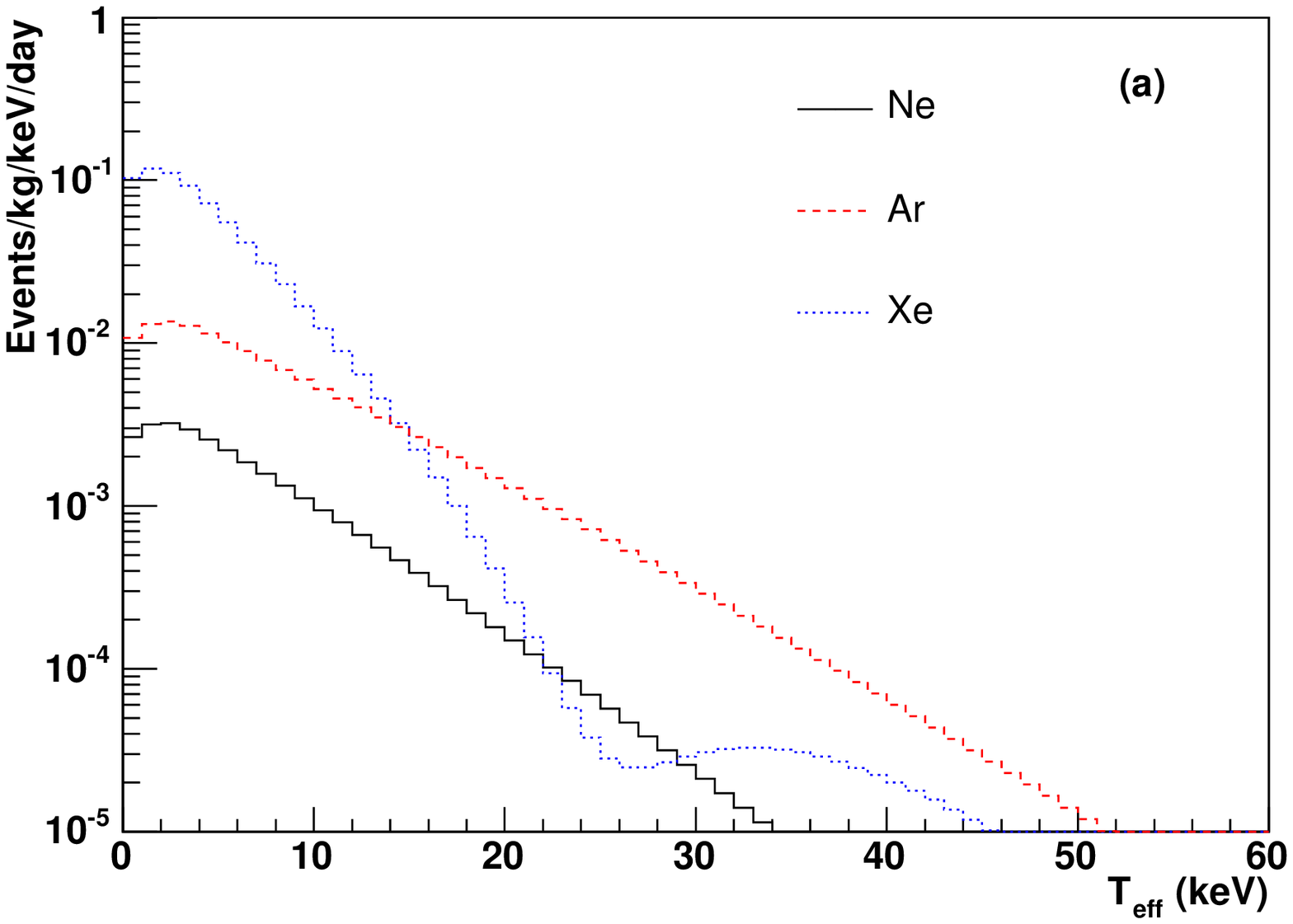}
\includegraphics[width=\figwidth]{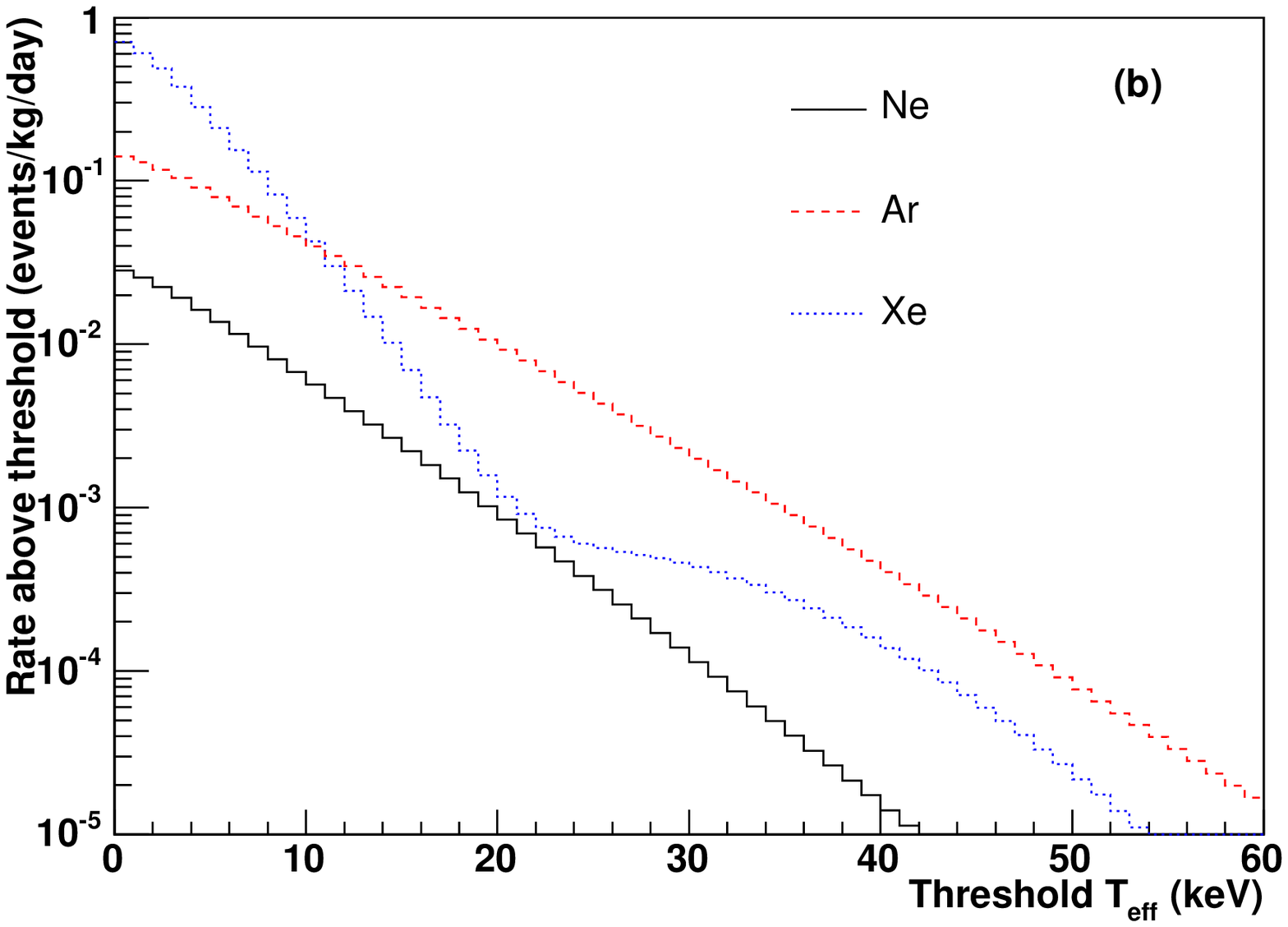}
\caption{\label{rates3}Predicted rates for WIMP scatters on Ne, Ar, and Xe assuming a WIMP-nucleon spin-independent cross-section $\sigma_{p} = 10^{-42}$ cm$^{2}$ and a 100-GeV WIMP mass. (a) Predicted rates in effective (observed) kinetic energy assuming 25$\%$ quenching for nuclear recoils.  (b) Integrated rates above threshold. }
\end{center}
\end{figure}

In general, a WIMP particle ($\chi$) can be detected through elastic scattering from a target nucleus (N);
\begin{equation}
\chi + \mbox{N} \longrightarrow \chi^{'} + \mbox{N}^{'},
\end{equation}
\noindent by measuring the kinetic energy imparted to the recoiling nucleus (N$^{'}$). The distributions of observed kinetic energies from WIMP scatters on neon, argon, and xenon are shown in Fig.~\ref{rates3}, for a WIMP-nucleon spin-independent cross-section of $10^{-42}$cm$^{2}$, a 100-GeV WIMP mass, and with the standard assumptions for 
the distribution of WIMPs in our galactic halo~\cite{jungman,lewin}. It is interesting to note that the integral rates in argon are equal to those in xenon for a threshold near $\sim10$ keV, due to the increased form factor suppression in xenon at higher nuclear recoil energies. If the material in which the nucleus recoils is a scintillator, the generated scintillation photons can be used to measure this energy. 
Shown in Table~\ref{scint_pars} are the scintillation parameters for liquid neon, argon, and xenon (see references~\cite{hitachi}-\nocite{neon_scint}\nocite{lanou_rayleigh}\nocite{argon_scint}\nocite{xenon_scint}\cite{neon_spectrum}) for nuclear recoil events (inferred from measurements with fission fragments) and for electrons.  Experimental determination of the scintillation parameters in liquid neon is currently being performed in support of the CLEAN~\cite{clean_preprint} experiment.

In general, ionizing radiation in noble liquids
will lead to the formation of dimers in either singlet or triplet states.  These states will have characteristic lifetimes ($\tau_{1}$ and $\tau_{3}$)
and are produced with different amplitudes ($I_{1}$ and $I_{3}$), depending on the liquid and the ionizing radiation.  Here we exploit these differences to separate electron from nuclear recoil events in liquid argon.
\begin{table*}
\caption{\label{scint_pars}Scintillation parameters for electrons and nuclear recoils in noble liquids from references~\cite{hitachi}-\cite{neon_spectrum}.  Quenching for nuclear recoils has been assumed to be 25\% for all materials. The ratio of prompt to late photons ($I_{1}/I_{3}$) provides the background discrimination.}
\begin{tabular}{llll } 
Parameter                 & Ne       & Ar     & Xe \\ \hline
Yield ($\times 10^{4}$ photons/MeV)       & 1.5    & 4.0  & 4.2 \\
prompt time constant $\tau_{1}$ (ns)           & 2.2      & 6      & 2.2   \\
late time constant   $\tau_{3}$            & $2.9~\mu$s      & $1.59~\mu$s   & 21 ns     \\
$I_{1}/I_{3}$ for electrons&          & 0.3    & 0.3 \\
$I_{1}/I_{3}$ for nuclear recoils&         & 3      & 1.6  \\
$\lambda(\mbox{peak})$ (nm)& 77      & 128    & 174 \\ 
Rayleigh scattering length (cm)      & 60     & 90     &  30  \\ \hline
\end{tabular}
\end{table*}

\begin{figure}
\begin{center}
\includegraphics[width=\figwidth]{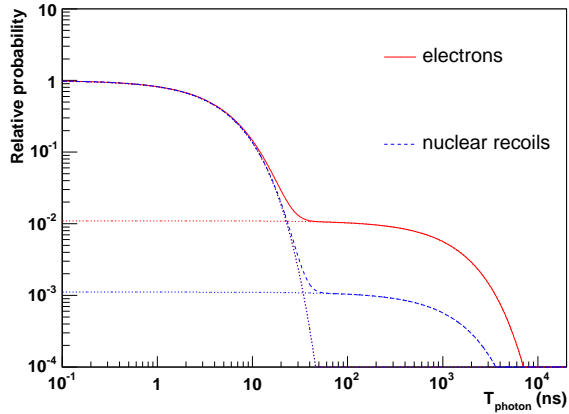}
\caption{\label{scinttime}Scintillation time distribution for electrons and nuclear recoils in liquid argon from Reference~\cite{hitachi}.  The dotted lines show the contributions from the singlet (prompt) and triplet (late) components separately.}
\end{center}
\end{figure}

\begin{figure}
\begin{center}
\includegraphics[width=\figwidth]{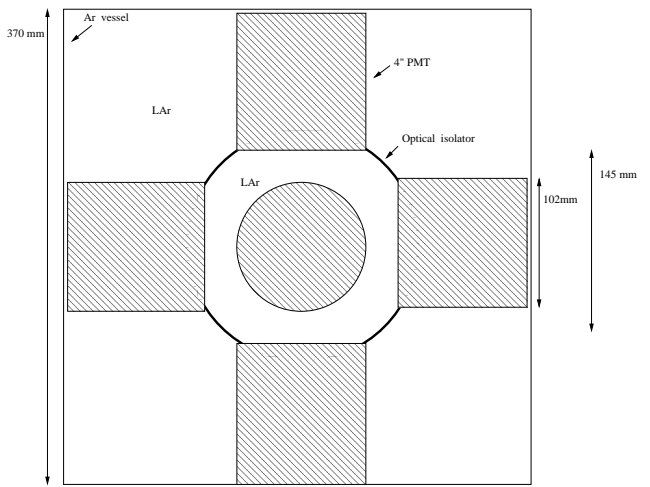}
\caption{\label{concept6}Conceptual design of 2 kg liquid argon experiment. 6 PMTs immersed in argon provide a photocathode coverage of approximately 75$\%$.  The PMT faces are coated with wavelength shifter to provide sensitivity to the 128 nm scintillation photons.}
\end{center}
\end{figure}

\begin{figure}
\begin{center}
\includegraphics[width=\figwidth]{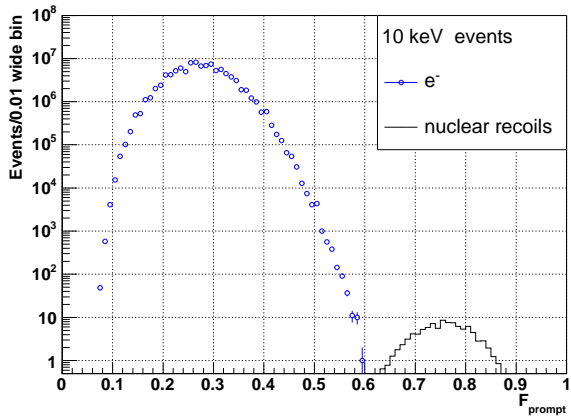}
\caption{\label{discrim}Discrimination of nuclear recoils from electrons in liquid argon.  Shown are the F$_{prompt}$ distributions from $10^{8}$ simulated electron events and for 100 simulated nuclear recoil events, with effective energies of 10 keV.}
\end{center}
\end{figure}

To estimate experimental sensitivity, a detector with a photosensitive coverage of 75\% has been assumed.  Nominal photomultiplier tube (PMT) properties assumed
are shown in Table~\ref{pmt_properties}, which have been extrapolated from properties of  PMTs used in the XENON~\cite{XENON} experiment.
The conceptual experimental design is shown in Fig.~\ref{concept6}.  Six PMTs, each with a 4 inch sensitive diameter are immersed in liquid argon, and view an inner
sensitive region, which is opaqued from activity in the outer region using an optical isolator.  Each PMT is coated with a wavelength shifting film to shift
the 128 nm photons into a region where the PMTs are sensitive.  It is assumed that the PMT pulses are digitized so that that the photon arrival times
are known to within the time resolution of the PMTs.  The ICARUS collaboration has constructed a large~(600 tonne)~liquid argon experiment, and the PMTs and wavelength shifter have been demonstrated to work in liquid argon~\cite{icarus}.  In this study the optical isolator is assumed to be a simple photon absorbing surface;
it may be advantageous to include a reflective foil coated with  wavelength shifter to 
increase the detector acceptance and photon yield, or equivalently to decrease the required number of PMTs.
  The time distributions for events with reconstructed energies of 10 keV have been simulated by sampling the scintillation
distributions for liquid argon (shown in Fig.~\ref{scinttime}) and applying the detection response of Table~\ref{pmt_properties}.     We have estimated the sensitivity here assuming a photon yield of 6 photoelectrons per keV in liquid argon, corresponding to a photocathode coverage of $75\%$ with a $20\%$ photon detection effiency.  From the simulated time distributions, the ratios
of prompt to total hits (F$_{prompt}$) have been constructed for electron and nuclear recoil events by taking the ratio of the number of hits in a prompt 100 ns window to those in a 15 $\mu$s window.  A simulated hardware trigger was used to determine the start time of the scintillation pulse, which is found by determining the time at which the number of hits in a sliding 100 ns window is maximized.  The trigger details would be optimized in an actual experiment with
calibration sources.    The resulting F$_{prompt}$ distributions are shown in Fig.~\ref{discrim}.  We find that there
is an essentially background-free region where the nuclear recoil events are expected, even in the presence of $10^{8}$ electron events.

\begin{figure}
\begin{center}
\includegraphics[width=\figwidth]{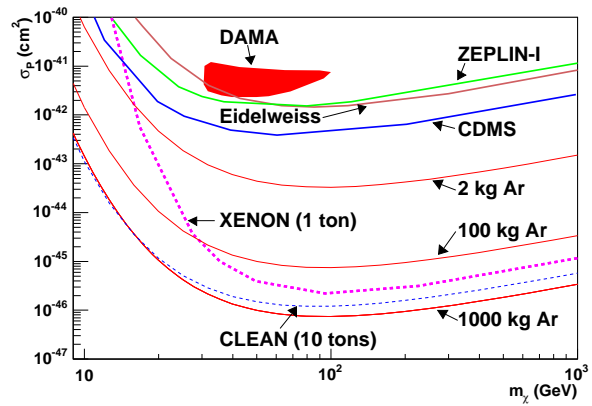}
\caption{\label{wimp_sens}Projected sensitivity to the WIMP-nucleon spin-independent cross-section versus WIMP mass.  A 2 kg argon experiment would improve current experimental limits, and could be scaled to large target masses. The limits for argon are for a one year background-free exposure, and a 10 keV threshold. Projected sensitivity for CLEAN is from Ref.~\cite{neon_sim}.  Data from other experiments adapted from Gaitskell and Mandic~\cite{dmtools}.}
\end{center}
\end{figure}

The largest source of background expected in liquid argon is from ${}^{39}$Ar, produced through cosmogenic activity in the atmosphere.
The ratio of ${}^{39}$Ar to natural argon is $8.1 \pm 0.3 \times 10^{-16}$~\cite{ar39loosli}.
The concentration of ${}^{42}$Ar in the atmosphere is much lower, $6 \times 10^{-21}$~\cite{ashitkov}, and it is not expected to be a significant source of background.
${}^{39}$Ar is a $\beta$ emitter with a Q-value of 565 keV, and a lifetime of 388 years~\cite{toi}.
The results presented here assume an analysis window between 10 and 25 keV, where the threshold energy of 10 keV would correspond to 60 PMT hits.
  The number of background events in this window for 2 kg of liquid argon, and from the 6 PMTs is shown in Table~\ref{pmt_back}.  The discrimination shown in Fig.~\ref{discrim}
is thus adequate to mitigate this level of background, and may be feasible for an experiment with several orders of magnitude larger mass.  We note the difference from the approach presented in Ref.~\cite{neon_sim} where 
a fiducial volume is defined to remove external-source background events.  In the approach presented here all of
the volume viewed by the PMTs is fiducial.
The 2 kg experiment could also be useful as a prototype for the design of a larger scale neon experiment (such as CLEAN),
where similar scintillation time discrimination could be used in a dual-purpose experiment with WIMP and solar neutrino sensitivity~\cite{neon_sim}.
The ultimate sensitivities to the spin-independent WIMP-nucleon cross-section for 2 kg, 100 kg, and 1000 kg experiments using liquid argon
are shown in Fig.~\ref{wimp_sens}.  It is found
that the 2 kg experiment has a a good sensitivity to the WIMP-nucleon cross-section, and larger scale argon experiments could increase the sensitivity
significantly.

\begin{table}
\caption{\label{pmt_properties}PMT and wavelength shifter (WS) properties assumed in the simulation.}
\begin{tabular}{ll} \hline
PMT diameter & 4 inches \\
U content    & 28 mBq   \\
Th content   & 16 mBq   \\
K  content   & 240 mBq   \\
PMT efficiency & 20 \%     \\ 
PMT time resolution & 5 ns     \\
PMT noise rate  & 1500 Hz \\
WS efficiency (4$\pi$) & 1 \\
WS time constant & 15 ns \\ \hline
\end{tabular}
\end{table}

The PMT background rates are assumed to be equivalent to the those in the R9288 PMT manufactured by Hamamatsu, scaled to 4 inches in diameter, and are  shown in Table~\ref{pmt_back}.  The experiment could equivalently use a larger number of the 2 inch diameter R9288 PMTs, and Hamamatsu has agreed to
provide a version of this PMT with a platinum underlay which should allow the PMT to function at liquid argon temperature~\cite{hamamatsu_private}.  With these assumed background
rates, the total neutron background from ($\alpha$,n) reactions in the PMTs is expected to be negligible~\cite{xenon_background}.  Other potential sources of neutron backgrounds to the WIMP search lead to similar requirements for cosmic-ray and neutron shielding, and materials selection as for a xenon dark matter experiment, described in reference~\cite{xenon_background}.

\begin{table}
\caption{\label{ar_experiments}Detector configurations and projected sensitivity to the WIMP-nucleon spin-independent cross-section ($\sigma_{p}$) with liquid argon for a one year exposure..}
\begin{tabular}{llll} \hline
Sensitive mass & Detector radius & No. PMTs & $\sigma_{p}$ sensitivity \\
(kg)           & (cm)            &          & (cm$^{2}$) \\ \hline
2 kg           &      7.25                 &   6       &  $\sim 10^{-43}$                         \\
100 kg         &      26                 & 79          &   $\sim 10^{-45}$                        \\ 
1000 kg        &       56               &  365        &    $\sim 10^{-46}$                     \\ \hline
\end{tabular}
\end{table}

\begin{table}
\caption{\label{pmt_back}Estimated backgrounds for the 2 kg, 6 PMT liquid argon experiment.  The dominant source
of background is from decays of ${}^{39}$Ar.  Shown are the estimated event rates per year in  the 10-25 keV analysis window.}
\begin{tabular}{ll} \hline
Source                 & Rate (events/year) \\ \hline
${}^{39}$Ar            &    1.26$\times 10^{6}$  \\
${}^{42}$Ar            &    76                   \\
U in PMTs              &    2.65$\times 10^{5}$  \\
Th in PMTs             &    1.51$\times 10^{5}$  \\
K in PMTs              &    4.54$\times 10^{5}$  \\ \hline
Total           &    2.13$\times 10^{6}$  \\ \hline
\end{tabular}
\end{table}

In summary, we have explored the possibility of using the differences in scintillation photon times for electrons and nuclear recoils in liquid argon for discrimination of backgrounds in a WIMP search. The level of discrimination should allow suppression of electron and $\gamma$-ray backgrounds required for a sensitive WIMP search.  A small-scale experiment using liquid argon, which would demonstrate this discrimination and lead to improved WIMP
sensitivity is presented.  This technique could be extended to a tonne-scale experiment, which would be competitive with a xenon-based experiment, but with a conceptually simpler design.

This work was supported with funds from the Los Alamos Directed Research and Development (LDRD) program.

\bibliography{ar}
\end{document}